# Multidimensional counting grids:
# Inferring word order from disordered bags of words

**Nebojsa Jojic**
Microsoft Research
Redmond, WA

**Alessandro Perina**
Microsoft Research
Redmond, WA

## Abstract

Models of bags of words typically assume topic mixing so that the words in a single bag come from a limited number of topics. We show here that many sets of bag of words exhibit a very different pattern of variation than the patterns that are efficiently captured by topic mixing. In many cases, from one bag of words to the next, the words disappear and new ones appear as if the theme slowly and smoothly shifted across documents (providing that the documents are somehow ordered). Examples of latent structure that describe such ordering are easily imagined. For example, the advancement of the date of the news stories is reflected in a smooth change over the theme of the day as certain evolving news stories fall out of favor and new events create new stories. Overlaps among the stories of consecutive days can be modeled by using windows over linearly arranged tight distributions over words. We show here that such strategy can be extended to multiple dimensions and cases where the ordering of data is not readily obvious. We demonstrate that this way of modeling covariation in word occurrences outperforms standard topic models in classification and prediction tasks in applications in biology, text modeling and computer vision.

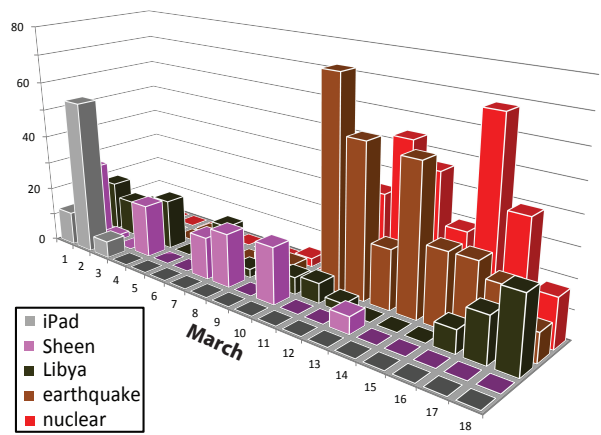

Figure 1: Usage counts for five terms in CNN news blog over the first 18 days of March

## 1  Introduction

In machine learning research, data samples are often represented as bags of words without particular order. This choice is often motivated by the difficulty or computational efficiency of modeling the known structure of the data, e.g., language. A striking example is the current computer vision research, where spatial structure of visual features is largely discarded by most object recognition algorithms. There are also examples of data where the structure is truly unknown. A gene expression array can be modeled as a bag of genes with expression levels simply corresponding to counts, because most of the time little is known about the cellular pathways that employ these genes. Without such knowledge there is no clear gene ordering. But biology is also abundant with situations where the raw data of interest truly has no structure. For example, the mammalian immune systems sees the virus inside the cell not as a whole but as a set of disordered peptides sampled inside from the viral proteins and presented on cellular surface for immune surveillance.

Clustering and dimensionality reduction techniques such as latent semantic indexing (LSI, (Deerwester et al., 1990)) and latent Dirichlet allocation (LDA, (Blei et al., 2003)) are among the most popular approaches to modeling disordered bags of words. In case of subspace models such as LDA, each bag of words is modeled as a mixture of topics, and each topic is represented by a tight distribution over words. In this paper, we point out that much of the variability in many interesting datasets is better modeled in terms of multidimensional thematic shifts, rather than outright mixing. In our model, certain words/features are dropped and others added as a consequence of movement in some hypothetical space. While our goal is to infer the properties of this space and the embedding of the data into it, it is useful to first consider an example of the data where such spatial embedding is directly available. Figure 1 shows the count of five different words in news stories in CNN's news blog[1] for the first 18 days of March 2011. A text document is often modeled as a bag of its words, and if the CNN's daily news blogs were to be modeled that way, then the highlighted words would participate in the bags with frequencies that indicate thematic shifts induced by the timeline of key events. The beginning of the month saw a spike in news on the launch of Apple's iPad 2, but words *Sheen*, and *Libya*, are also abundant. A week later, it seems that the interest of the American public quickly shifted to the events in the professional and personal life of actor Charlie Sheen, overshadowing a steady trickle of news on the uprising in the Middle East. Then the catastrophic natural events in Japan caused a large spike of the usage of the word *earthquake*, followed a few days later by a significant increase in the usage of the word *nuclear*, reflecting the problems in the nuclear plants that started to unfold as consequence of the earthquake and the ensuing tsunami. With the expectation of the UN resolution on Libya, this word regains in its dominance towards the end of this period. From a signal processing perspective, the distribution of the word counts across the timeline seems to be caused by a few point sources, with this excitation then going through an averaging filter. News people are used to the thematic shifts of this nature, where each new story enjoys some limited life time and is then suppressed by other more interesting topics. From the machine learning perspective, this situation can be well modeled by creating a series of relatively tight word distributions (corresponding to point sources on the timeline), and then combining several consecutive distributions to form the expected histogram of words for any given day. The thematic shifts over the days are then simply modeled by moving the averaging window across the timeline of these point sources. Thus even though each blog is considered a disordered bag of words, a good model would in fact order them along a line to induce constraints on the word mixing that gives rise to the observed bags of words.

However, even in case of data where the additional metadata that would provide such ordering (date in this case) does not exist, we can see thematic shifts.

---

[1] http://news.blogs.cnn.com/

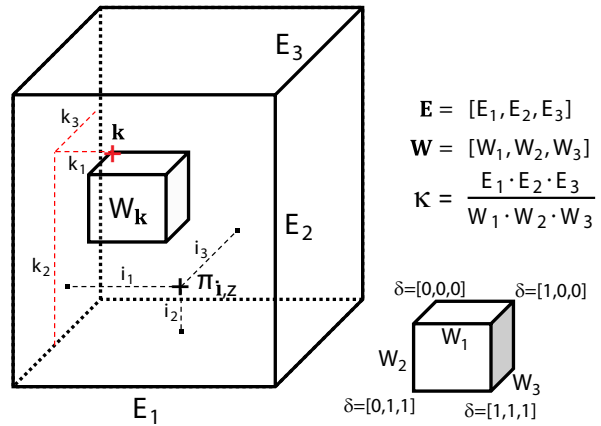

Figure 2: An example of a counting grid geometry. In general, the data is embedded into a hypercube which is wrapped around along each dimension to avoid local minima that would be caused by abrupt cuts along any dimension.

In this paper we provide a simple model for data in which much of the variation is induced by thematic shifts expressed as movement through an inferred space. As in the news example, we assume that there is some space into which a set of tight distributions is embedded, and that these distributions are then combined using a windowing operation to create a resultant distribution from which the observed bags of words or features are generated. However, we do not assume that the mapping is given a priori. For simplicity, we assume that the space is a discrete grid of counts, but of arbitrary dimension (we experimented with 2- and 3-dimensional grids) and we consider iterative estimation of counts on this grid and the mapping of the data to the overlapping windows on it. Our experiments indicate that the thematic shifts are indeed present in a variety of datasets, and as a result, our model outperforms standard topic mixing (LDA) there. We analyzed a wide variety of data types, including text, images, gene expression and viral peptides, and used the learned counting grids to perform regression or classification.

## 2 The counting grid model

Formally, the basic counting grid $\pi_{\mathbf{i},z}$ is a set of normalized counts of words/features indexed by $z$ on the $D$-dimensional discrete grid indexed by $\mathbf{i} = (i_1, \ldots, i_D)$ where each $i_d \in [1 \ldots E_d]$ and $\mathbf{E} = (E_1, \ldots, E_D)$ describes the extent of the counting grid. Since $\pi$ is a grid of distributions, $\sum_z \pi_{\mathbf{i},z} = 1$ everywhere on the grid. A given bag of words/features, represented by counts $\{c_z\}$ is assumed to follow a count distribution found somewhere in the counting grid. In particular, using windows of dimensions $\mathbf{W} = [W_1, \ldots, W_D]$, each bag can be generated by first averaging all counts in the hypercube window $W_\mathbf{k} = [\mathbf{k} \ldots \mathbf{k} + \mathbf{W}]$ starting at $D$-dimensional grid location $\mathbf{k}$ and extending in each direction $d$ by $W_d$ grid positions to form the histogram $h_{\mathbf{k},z} = \frac{1}{\prod_d W_d} \sum_{\mathbf{i} \in W_\mathbf{k}} \pi_{\mathbf{i},z}$, and then generating a set of features in the bag. In other words, the position of the window $\mathbf{k}$ in the grid is a latent variable given which the probability of the bag of features $\{c_z\}$ is

$$p(\{c_z\}|\mathbf{k}) = \prod_z (h_{\mathbf{k},z})^{c_z} = \frac{1}{\prod_d W_d} \prod_z \Big( \sum_{\mathbf{i} \in W_\mathbf{k}} \pi_{\mathbf{i},z} \Big)^{c_z}, \quad (1)$$

Relaxing the terminology, we will refer to $\mathbf{E}$ and $\mathbf{W}$ respectively as the counting grid and the window size. We will also often refer to the ratio of the window volumes, $\kappa$, as a capacity of the model in terms of an *equivalent number of topics*, as this is how many nonoverlapping windows can be fit onto the grid. Fine variation achievable by moving the windows in between any two close by but nonoverlapping windows is useful if we expect such smooth thematic shifts to occur in the data, and we illustrate in our experiments that indeed it does. Finally, with $W_\mathbf{k}$ we indicate the particular window placed at location $\mathbf{k}$ (see Fig. 2).

### 2.1 Inference and learning

To compute the log likelihood of the data, $\log P$, we need to sum over the latent variables $\mathbf{k}$ before computing the logarithm, which, as in mixture models, or as in epitomes (Jojic et al., 2003), which are much more similar to the counting grids, makes it difficult to perform assignment of the latent variables (in our case positions in the counting grid) while also estimating the model parameters. This makes an iterative exact or a variational EM algorithm necessary. Bounding (variationally), the non-constant part of $\log P$, we get

$$\log P \geq B = - \sum_t \sum_{\mathbf{k}^t} q_{\mathbf{k}^t} \log q_{\mathbf{k}^t} + \quad (2)$$
$$+ \sum_t \sum_{\mathbf{k}^t} q_{\mathbf{k}^t} \sum_z c_z^t \log \sum_{\mathbf{i} \in W_{\mathbf{k}^t}} \pi_{\mathbf{i},z},$$

where $q_{\mathbf{k}^t}$, or in shorthand, $q_\mathbf{k}^t$ is the variational distribution over the latent mapping onto the counting grid of the $t$-th bag. Each of these variational distributions can be varied to maximize the bound. In fact, for a given counting grid $\pi$, the bound is maximized when each distribution $q^t$ is equal to the exact posterior distribution. This is a standard variational derivation of the exact E step, which leads to

$$q_\mathbf{i}^t \propto \exp \sum_z c_z^t \cdot \log h_{\mathbf{i},z}, \quad (3)$$

which simply establishes that the choice of $\mathbf{k}$ should minimize the KL divergence between the counts in the bag and the counts $h_{\mathbf{k},z} = \sum \pi_{\mathbf{i},z}$ in the appropriate window $W_\mathbf{k}$ in the counting grid. For each $t$, the above expression is normalized over all possible window choices $\mathbf{k}$.

To optimize the bound $B$ with respect to parameters we note first that it is the second term in Eq. 1 that involves these parameters, and that it requires another summation before applying the logarithm. The summation is over the grid positions $\mathbf{i}$ within the window $W_\mathbf{k}$, which we can again bound using a variational distribution and the Jensen's inequality:

$$\log \sum_{\mathbf{i} \in W_{\mathbf{k}^t}} \pi_{\mathbf{i},z} = \log \sum_{\mathbf{i} \in W_{\mathbf{k}^t}} r_{\mathbf{i},\mathbf{k},z}^t \cdot \frac{\pi_{\mathbf{i},z}}{r_{\mathbf{i},\mathbf{k},z}^t} \geq \sum_{\mathbf{i} \in W_{\mathbf{k}^t}} r_{\mathbf{i},\mathbf{k},z}^t \log \frac{\pi_{\mathbf{i},z}}{r_{\mathbf{i},\mathbf{k},z}^t} \quad (4)$$

where $r_{\mathbf{i},\mathbf{k},z}^t$ is a distribution over locations $\mathbf{i}$, i.e. $r$ is positive and $\sum_{\mathbf{i} \in W_\mathbf{k}} r_{\mathbf{i},\mathbf{k},z}^t = 1$ and is indexed by $\mathbf{k}$ as the normalization is done differently in each window, and is indexed by $z$ as it can be different for different features, and indexed by $t$ as the term is inside the summation over $t$, so a different distribution $r$ could be needed for different bags $\{c_z^t\}$. This distribution could be thought of as information about what proportion of these $c_z$ features of type $z$ was contributed by each of the different sources $\pi_{\mathbf{i},z}$ in the window $W_\mathbf{k}$. However, by performing constrained optimization (so that $r$ adds up to one), we find that assuming a fixed set of parameters $\pi$, the distribution $r_{\mathbf{i},\mathbf{k},z}^t$ that maximizes the bound is independent of $t$, i.e., the same for each bag:

$$r_{\mathbf{i},\mathbf{k},z}^t = \frac{\pi_{\mathbf{i},z}}{\sum_{\mathbf{i} \in W_\mathbf{k}} \pi_{\mathbf{i},z}} = \frac{\pi_{\mathbf{i},z}}{\prod_d W_d \cdot h_{\mathbf{k},z}}. \quad (5)$$

If we do consider distributions $r$ as a feature mapping to the counting grid, then this result is again intuitive. If all we know is that a bag containing $c_z$ features of type $z$ is mapped to the grid section $W_\mathbf{k}$, and have no additional information about what proportions of these $c_z$ features were contributed from different incremental counts $\pi_{\mathbf{i},z}$, then the best guess is that these proportions follow the proportions among $\pi_{\mathbf{i},z}$ inside the window. If we assume now that $r$ and $q$ distributions are fixed, then combining Eqs. 3,4 and minimizing the resulting bound wrt parameters $\pi_{\mathbf{i},z}$ under the normalization constraint over features $z$, we obtain the update rule,

$$\hat{\pi}_{\mathbf{i},z} \propto \sum_t \sum_{\mathbf{k} | \mathbf{i} \in W_\mathbf{k}} q_\mathbf{k}^t \cdot c_z^t \cdot r_{\mathbf{i},\mathbf{k},z}^t \quad (6)$$

which by Eq. 5 reduces to

$$\hat{\pi}_{\mathbf{i},z} \propto \pi_{\mathbf{i},z} \cdot \sum_t c_z^t \sum_{\mathbf{k} | \mathbf{i} \in W_\mathbf{k}} \frac{q_\mathbf{k}^t}{h_{\mathbf{k},z}} \quad (7)$$

The steps in Eqs. 3 and 7 constitute the E and M step which can be iterated till convergence (within a desired precision). The first step aligns all bags of features to grid windows that (re)match the bags' histograms, and the second re-estimates the counting grid so that these same histogram matches are even better. Thus, starting with non-informative (but symmetry breaking) initialization, this iterative process will jointly estimate the counting grid and align all bags to it. To avoid severe local minima, it is important, however, to consider the **counting grid as a torus**, and consider all windowing operations accordingly, as was previously proposed for learning epitomes (Jojic et al., 2003, 2010), a model that quilts spatially-organized images and videos. This prevents the problems with grid boundaries which otherwise could not be crossed when more space is needed to grow the layout of the features.

### 2.2 Computational efficiency

Careful examination of the steps reveals that by the efficient use of cumulative sums, both the E and M steps are linear in the size of the counting grid. Both steps require computing $\sum_{\mathbf{i} \in W_\mathbf{k}} f_\mathbf{i}$, which can be done by first computing, in linear time the cumulative sums of $f$ and then computing appropriate linear combinations.

For example, in the 2D case we have $\mathbf{i} = (i,j)$, $\mathbf{k} = (k,\ell)$ and one can compute the cumulative sum $F_{m,n} = \sum_{(i,j) \leq (m,n)} f_{i,j}$, and then set $\sum_{(i,j) \in W_{k,\ell}} f_{i,j} = F_{k+W_1+1, \ell+W_2+1} - F_{k, \ell+W_2+1} - F_{k+W_1+1, \ell} + F_{k,\ell}$.

This can be generalized by associating to each vertex $v$ of the hypercube $W_\mathbf{k}$ ($2^D$ vertices in total) a binary vector $\delta^v$. Different vertices of $W_\mathbf{k}$ share various coordinates, as along a dimension, say $d$, a vertex $v$ can only assume two values ($k_d$ or $k_d + W_d$). We define elements of vector $\delta_v$ as follows

$$\delta_d^v = \begin{cases} 1 & \text{if } v_d = k_d + W_d \\ 0 & \text{else} \end{cases}$$

The value of $\delta$ for some vertex is shown in Fig.2. Given this we can write

$$\sum_{\mathbf{i} \in W_\mathbf{k}} f_\mathbf{i} = \sum_{v=1}^{2^D} (-1)^{|1-\delta^v|} \cdot F_{\mathbf{i}+\delta^v \circ \mathbf{W}} \quad (8)$$

where the ∘ is the pointwise multiplication.

### 2.3 Label embedding

Once a CG is learned, one may embed *other* discrete (e.g., class labels $y^t = l$, $l = 1...L$) or continuous (e.g., HIV $y^t$) values on the grid. This is achieved using the posterior probabilities $q_\mathbf{k}^t$ already inferred and simply computing the M-step (Eq. 7) using the target label in place of counts ($c_z^t$):

$$\gamma(\mathbf{i}, l) = \frac{\sum_t \sum_{\mathbf{k}|\mathbf{i} \in W_\mathbf{k}} q_\mathbf{k}^t \cdot [y^t = l]}{\sum_t \sum_{\mathbf{k}|\mathbf{i} \in W_\mathbf{k}} q_\mathbf{k}^t} \quad (9)$$

$$or \quad \gamma(\mathbf{i}) = \frac{\sum_t \sum_{\mathbf{k}|\mathbf{i} \in W_\mathbf{k}} q_\mathbf{k}^t \cdot y^t}{\sum_t \sum_{\mathbf{k}|\mathbf{i} \in W_\mathbf{k}} q_\mathbf{k}^t}. \quad (10)$$

where (9) is used for discrete, and (10) for the continuous labels.

The label embedding $\gamma$ can then be used for classification or regression, in what is essentially a nearest-neighbor strategy: When a new data point is embedded based on its bag of words, the label is simply read out from $\gamma$, which is dominated by the training points which were mapped in the same region. Other options to using the CGs for classification/regression are of course available, but we opted for this simplest one in our experiments, for the reasons we discuss next.

## 3 Experiments

As we are mostly interested in the quality of unsupervised learning of the distribution over the bags of words, in the majority of the experiments we compared LDA with CGs in the following setting. Each data sample consists of a bag of words and a label. The bags were used without labels to train a CG and LDA models that capture covariation in word occurrences, with CGs mostly modeling thematic shifts, and LDA modeling topic mixing. Then, the label prediction task is performed in a leave-one-out setting, where each data point is taken out of the set used to estimate the label embedding $\gamma$ of the rest of the data, and then the left out sample's mapping is used to read out the $\gamma$ in the appropriate location (or if the posterior is uncertain, to average out the embedded labels accordingly). LDA is used in a similar fashion: the most similarly mapped points are used to predict the label for the left out sample. For LDA we have also investigated linear regression based on the latent mapping. Finally, a variety of other methods are occasionally compared to, and slightly different evaluation methods described in individual subsections, when appropriate.

### 3.1 Text: CMU 20 Newsgroup

In the CMU newsgroup dataset[2], each news post is treated as a document (a bag of words) labeled by one of 20 labels representing the news group of its origin (as opposed to our introductory toy example, the date of the post was not available). Following previous work (Banerjee and Basu, 2007) we reduced the dataset into subsets with varying similarities among the news groups. We consider and report here the results only for the more challenging, **news-20-same**, with 1700 posts from the highly related groups `comp.os.ms-windows`, `comp.windows.x` and `comp.graphics`. We first compared the embedding of documents from all three classes provided by CGs (a torus with overlapping windows) and LDA (a simplex of topic proportions) based respectively on label embedding function $\gamma$, and on the performance of the K-Nearest Neighbor classifier (K=3) using the topic proportions $\theta$ of LDA employing the KL divergence to evaluate distances. As in (Reisinger et al., 2010), comparisons with LDA were performed in $10 \times 10$-fold crossvalidation. For both methods in each training/split test, we reduced the vocabulary to 2000 words using the feature reduction method of (Peng et al., 2005) on the training data to save on computation time. We have also evaluated the use of simple bag of words comparisons in kNN setting for classification. Figure 3a summarizes the CG results: We learned several 2D and 3D grids, varying **E**, **W** and $\kappa$. We repeated the tests trying to take the same $\kappa$ for the 2D and 3D cases to evaluate how document classes embed in spaces of different dimensions[3]. In all the tests, 3D counting grids have been found to perform slightly better suggesting that more than 2 dimensions are needed to embed such complex classes. For sake of comparison, we also added recent results from (Reisinger et al., 2010), including their test of vonMF model and their best result for their own new model, spherical admixture model (SAM), which models topics using vonMF distribution and tf-idf features as input. CG outperforms other models (and with an even larger margin if SAM uses the simple count (tf) features, as CGs did in our experiments rather than the more complex derived features as in the citation; For example, CGs reduce SAM's classification error in this setting by over 30%).

---

[2] http://www.cs.cmu.edu/afs/cs.cmu.edu/project/theo-20/www/data/news20.html

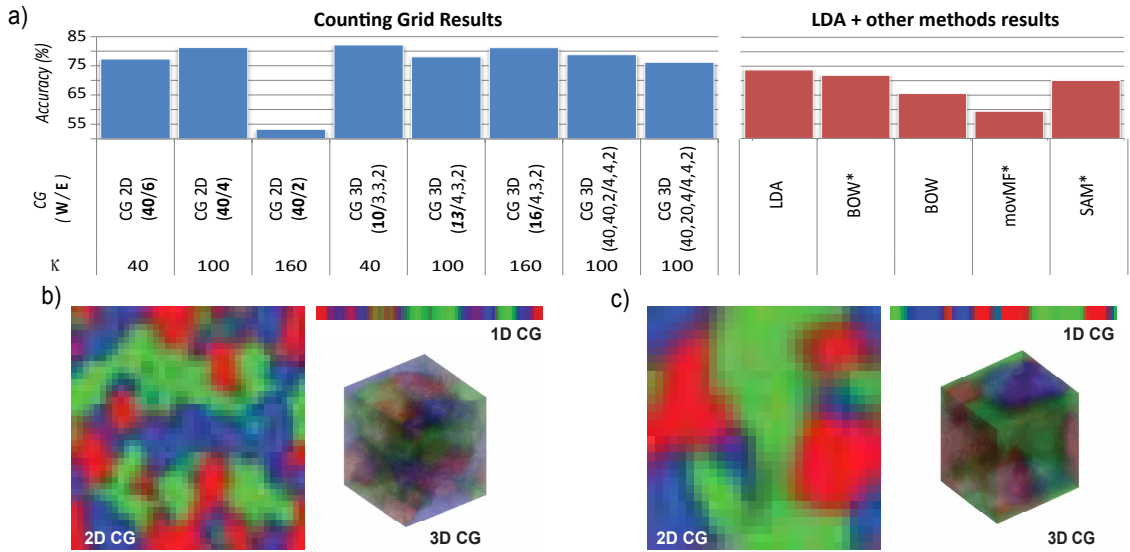

Figure 3: News-20 classification results. BoW stands for bag-of-words, movBF is the mixture of von-Mises Fisher (Mardia and Jupp, 2000). The asterisk *, indicates a method that uses the original feature set. We describe a particular CG by specifying **E** and **W** in the form **E/W**. If only one value is reported for E or W, then the CG has the same size in all the dimensions (hypercube). b) Embedding of the three class labels in 1D 2D and 3D CGs . ( The accuracy for 1D CGs does not exceed 60%). c) Embedding of the three class labels of the easiest subset of the 20 newsgroup dataset: **news-20-different**, with posts from `rec.sport.baseball`, `sci.space` and `alt.atheism` (see (Banerjee and Basu, 2007) for details).

To illustrate the label embedding $\gamma$ for the training documents in the learned counting grids, we show where the image labels for the three classes in each dataset ended up in one of the tests: the CG is colored red green and blue based on the fractions of the documents from each of the three classes mapped to the CG locations (Fig. 3b ). The 3D grids are rendered semi-transparent. The figure demonstrates a complex structure with lots of transitions among classes in 2D.

---
[3]For example, the 2D grid E = [40,40], W = [4,4], has roughly the same $\kappa$ of the 3D grid of size E=[10,10,10], W=[3,3,2]

For comparison, in (Fig. 3c), we show the embedding for an easier subset of the newsgroup dataset where the blobs of data points of the same class are larger and better separated. The best capacity was found to be $\kappa \approx 100$ for both 2 and 3 dimensions. Interestingly, keeping the same ratio, but varying the dimensions of the window we find that small windows (i.e., size $2 \times 2$) provides gradation of window overlap that is too coarse (either half the window, or full overlap, or no overlap), while increasing the size of the window beyond 4 along each dimension to allow for more refined overlaps, did not increase the performance. It is interesting to see that the model does not easily over train. Although for example the $40 \times 40$ counting grid consists of 1600 independent word distributions $\pi_{\mathbf{i},z}$, these positive parameters are summed up in large groups to represent the data and these groups (windows) have a large overlap. And so, while 100 independent nonoverlapping windows for $\kappa = 100$ can be designated on the grid, cramming in more components expressible as sums of the subsets of the same set of positive numbers is hard unless some real structure in data is discovered. Only for small window sizes do we start to see overtraining. Additional results for the 2D case are reported in table 1 and show that CG are quite robust to the choice of the grid size, when given enough room ($\kappa$) to accommodate for the variation in the documents[4].

Table 1: Classification rate (%) vs CGs size for a fixed ratio $\kappa$ of grid size to window size. CGs are squared so we report only one dimensions, i.e., $\mathbf{E} = \mathbf{20} = [20, 20]$.

| E/ W | 20/2 | 40/4 | 60/6 | 80/8 | 100/10 |
|---|---|---|---|---|---|
| Acc. | 39.30 | 81.28 | 80.34 | 78.22 | 78.18 |

### 3.2 Images: Scene classification

In case of visual data, modeled as bags of features, CG model can account for misalignment of scenes as a change of location in the counting grid. The smooth thematic motion here corresponds to the motion of the camera. The visual scene dataset, introduced in (Oliva and Torralba, 2001), is composed by two datasets composed by four natural and four artificial (man-made) categories and it is widely used by the vision community. Each class contains roughly 250 images.
Following the standard bag-of-*visual*-words (Li and Perona, 2005) approach we extracted SIFT features from 16x16 pixel windows computed over a grid spaced of 8 pixels and clustered the descriptors in $Z = 200$ visual words. We describe an image as a bag of their features. The feature maps for images are originally of the size $30 \times 30$ feature maps. For the 2D case, we used this for the window size, as the original features really did have a spatial arrangement in a window this size. We kept the same volume for the 3D case, setting $W = [10, 10, 9]$. We varied the grid size, keeping cubic grids, choosing it form the set $[2, 3, 4]$. As discussed above, we ran the experiments to compare CG and LDA representations in terms of the power of the kNN classifier using their unsupervised data embedding, i.e., we used all training and test images of all classes to estimate a CG or an LDA thus embedding the data into the space of hidden variables for these models without using the scene labels, and then used these hidden variables (position in the grid for CG or topic proportions for LDA) to perform $10 \times 10$ cross-validation on label prediction in appropriate training/test splits. Results, including SAM (Reisinger et al., 2010) which was performed on the same data are shown in figure 4a as bars (stacked red+blue bars, labeled with "90%"). As in (Reisinger et al., 2010) we repeated the evaluation but inverting training and test sets so that training used only 1/10-th of the data (blue bars, labeled with "10%"). In both cases CGs outperform other methods and is robust to overtraining.

We also ran experiments we dubbed *supervised classification* in which we trained a model (CG or LDA) per class, varying the cardinality of the training set. Subsequently we classified the test samples using likelihood tests. We repeated this process 5 times, averaging the results. We compared 2D, 3D CGs and LDA in figure 4b, where the two dimensional case seems to outperform the higher dimensions (we also tried 4 dimensions) and in general, CGs outperform LDA with a large margin across the training data cardinalities.
In figure 4c we compared the 2D CGs with the similar approach of (Perina and Jojic, 2011), where in the M-step the original structure of the image is kept and placed onto the grid and where the window size is equal to the window size. Results show that even discarding all the spatial information, the accuracy does not drop much.
Finally we compared with the state of the art combining CGs with the discriminative layer put by (Perina et al., 2009). We reached the 91.12% and 91.77% in the 2D case, 91.01% and 89.49% in the 3D case, respectively on Natural and on Artificial images, matching with statistical significance the performance of the well known method presented in (Bosch et al., 2006) (Accuracies of 90.2% and 92.5%).

---
[4]We reached similar conclusions for the 3D case.

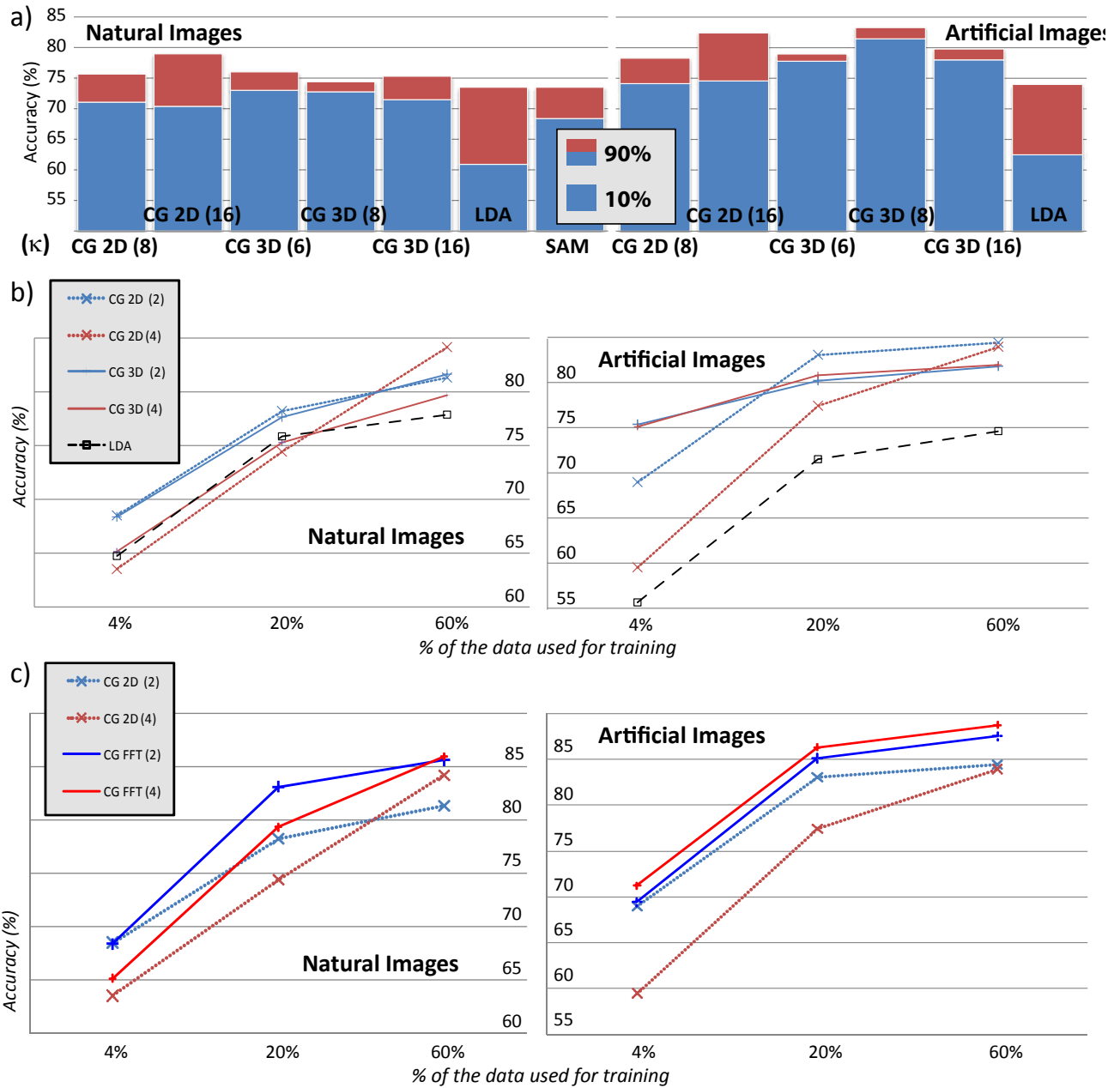

Figure 4: Scene classification. a) Unsupervised embedding evaluation in terms of the classification based on the latent space. (Reisinger et al., 2010) (SAM) does not report result for "Artificial" dataset. We kept fixed **W** and we varied **E**; The number in brackets is the coverage $\kappa$. b-c) Supervised image classification results using different counting grids and LDA models for different classes (each class contains 250 images). See text for details.

### 3.3 Biology: Cellular presentation of viral peptides and viral load

The immune system is among the most interesting and most complex adaptive systems in higher organisms. It consists of a number of interacting subsystems employing various infection clearing paths, and cellular presentation plays a central role in many of them. Most of the cells present a sample of peptides derived from cellular proteins as a means of advertising their states to the immune system. This facilitates globally coordinated action against viral infection. As the immune pressure depends on cellular presentation, the variation in cellular presentation across patients is expected to reflect on the variation in viral load, at least to some extent (Moore et al., 2002; Hertz et al., 2010). We analyzed predicted cellular presentation of 181 HIV patients from the Western Australia cohort (Moore et al., 2002). To avoid confounding effects of HIV clade, we analyzed only the clade B infected patients. We represented each patient's cellular presentation by a set of 492 counts over that many 9-long peptides from the Gag protein, previously found to be targeted by the immune system. The counts were calculated based on the patients HLA class I types and the HLA-peptide binding estimation procedure discussed in (Hertz et al., 2010). We trained counting grids and LDA models of varying complexity on all the data and then predicted the patients viral load based on the representation of the data in the latent space in leave-one-out cross validation framework, as discussed above (Fig. 5). We found that 3D counting grids outperformed slightly the 2D grids in this case, and that both outperformed LDA. In Fig. 6 we also show the evolution of the embedding $\gamma$ of the log viral loads for the first 180 patients (used to predict the log viral load of the 181st patient) over the iterations of counting grid learning. As discussed above, the bags of words (peptides) are mapped to the counting grid iteratively as the grid is estimated as to best model the bags, but the regression target, the viral load, was not used during the learning of CGs or LDA models. However, the inferred mapping after each iteration can be used to visualize how the embedded viral load $\gamma$ evolves. The emergence of areas of high (red) and low (blue) viral load indicates that as the structure in the cellular presentation is discovered, it does indeed reflect the variation in viral load. The presented correlation factors between true and predicted viral loads are computed after the convergence. The correlation factors of above 0.3 which we uniformly find across a range of values $\kappa$ indicate that cellular presentation of the Gag protein explains more than 9% of the log viral load. In comparison, targeting efficiency analysis of Gag (Hertz et al., 2010) could only explain less than 4% of viral load. Although viral load varies dramatically across patients for a variety of reasons, e.g. gender, previous exposures to related viruses, etc., detection of statistically significant links between cellular presentation and viral load is expected to have important consequences to vaccine research (Kiepiela et al., 2004) .

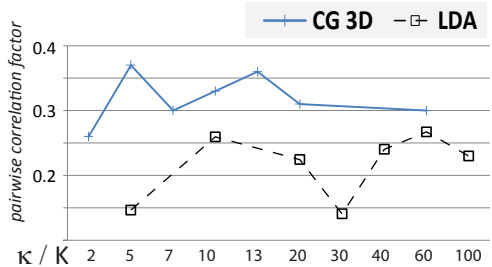

Figure 5: HIV viral load regression. Shown is the variation of the correlation factor $\rho$ for a range of capacities $\kappa$ and number of LDA topics K.

### 3.4 Biology: promoter classification

We considered a dataset composed by E. coli promoter gene sequences (DNA) with associated imperfect domain theory (Towell et al., 1990). The task is to recognize promoters in strings that represent nucleotides (A, G, T, or C). A promoter is a genetic region which initiates the first step in the expression of an adjacent gene (transcription). The input features are 57 sequential DNA nucleotides (fixed length). A special notation is used to simplify specifying locations in the DNA sequence. The biological literature counts locations relative to the site where transcription begins. Fifty nucleotides before and six following this location constitute an example. We transformed the sequences in bag of features representations as we explained in the previous section. Results, obtained using leave-one-out (LOO) validation, show that CGs ( 83.01%) outperform HMM ( 67.3% ) and methods based on them (Jaakkola and Haussler, 1998), as the fisher kernel ( 79.2% ) which are specialized for sequences.

### 3.5 Biology: Microarray expression classification

Previous work (Perina et al., 2010) has interpreted microarray expression values as counts in "bags-of-genes", with good classification rates have been reached. Following the same intuition we perform here microarray classification.

We used the dataset of the study of prostate cancer in (Dhanasekaran, 2001) consisting of 54 samples with 9984 features. The samples are subdivided in different classes: 14 samples are labelled as benign prostatic hyperplasia (labelled BPH), 3 as normal adjacent prostate (NAP), 1 as normal adjacent tumor (NAT), 14 as localized prostate cancer (PCA), 1 prostatitis (PRO) and 20 as metastatic tumors (MET). The 6 classes can be divided in three macro-classes: non-cancer (BPH,NAP,PRO), cancer (NAT,PCA), metastatic tumor (MET). As in (Rogers et al., 2005), we filtered the genes by variance and keep only the most variable five hundred.

We compare the results of 2D and 3D with LDA as we did in the previous experiments. Classification errors have been computed using 10-fold cross validation. We learned several *squared* counting grids of several capacities by picking different values for the width of the square 2D window from the set $\{5, 8, 11\}$, and simi-larly the width of the square 2D counting grid from $\{23, 32, 42, 57\}$, and by picking the width of the cubic 3D window from $\{3, 4, 5\}$ , and the cubic 3D counting grid width from $\{8, 10, 12, 14\}$. We tried to keep capacity factor $\kappa$ comparable between the 2D and the 3D cases. We compared our results with the LDA (Rogers et al., 2005), varying the number of topics K, in the set of $\kappa$. Results shown in Fig.7 demonstrate that CGs outperform with LDA across a range of choices of K. It is worth noting that to the best of our knowledge, the state-of-the art on this dataset is 91.2% (Perina et al., 2010) and 2D grids outperform this value across a range of complexities. We also run classification tasks on the brain and colon cancer datasets (Pomeroy, 2002; Alon et al., 1999) which consist of 5 and 2 classes, respectively. In this case we fixed **W** and **E** to best values we found for the previous microarray set, and we embed the data for this single complexity and then performed classification as above. The results are summarized in Tab. 2 and compared with the LDA results published in (Perina et al., 2010). In this cases, 3D CGs performs better and the result we got on Colon cancer dataset is currently the state of the art.

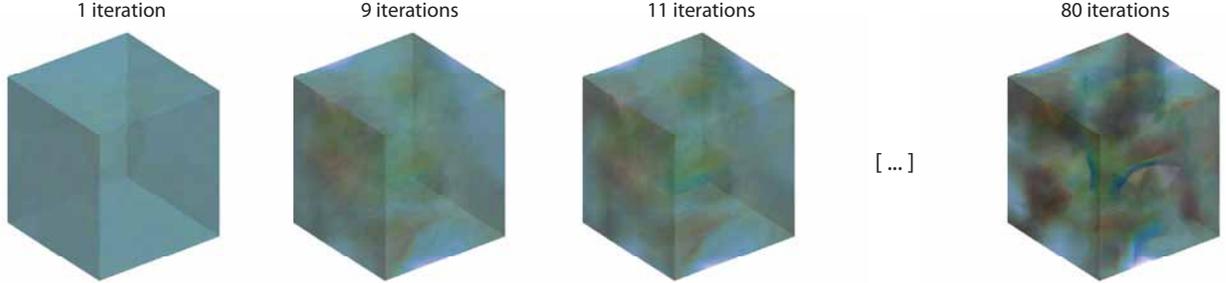

Figure 6: HIV viral load embedding.

Table 2: Microarray classification results (%).

| Dataset | CG 2D | CG 3D | LDA |
|---|---|---|---|
| Brain (Pomeroy, 2002) | 86.30 | 87.84 | 82.11 |
| Colon (Alon et al., 1999) | 87.40 | 89.20 | 76.62 |

## 4 Conclusions

We have introduced the multidimensional counting grid model which outperforms the subspace models in modeling a variety of datasets where we found that thematic shifts seem to be a better fit to capturing correlations in word occurrence. We found that 2D grids outperform 3D grids in the visual scene classification for most scenes, which indicates the possibility that the model primarily captures image mis-alignment. On the other hand, the topology needed to capture variation in cellular presentation of HIV Gag is better embedded in 3D counting grid. In fact, 3D counting grids are slightly better in most, but not all applications, and more research is needed into the effect that dimensionality as well as aspect ratio of the counting grids have on model quality in various applications. Another interesting observation that warrants more research is the fact that while the optimal window size for text seemed to be relatively small (up to 4 in one dimension), the HIV presentation modeling required larger patches than text modeling (using 8×8×8 windows into a larger grid perform slightly better than 4 × 4 × 4 windowing of smaller grids with the same ratio $\kappa$. Larger patch sizes allow for finer-grained overlapping of the data on the counting grid. Finally, the LDA and CG models seem to model slightly different aspects of the data, and while LDA may suffer from having to model thematic shifts in data that has them, the basic CG model does not allow for other types of mixing (In images for example, CGs capture camera motion across scenes, but LDA might capture mixing of different objects, and we can imagine this analogy carried over to other types of data). Thus, a straightforward ways of combining aspects of both models should be investigated further. The most similar previously published model to counting grids is the epitome model (Jojic et al., 2003, 2010; Ni et al., 2008), which was also based on overlapping patches in the latent space, and which also reaped benefits from shift-invariance. However, epitomes relied on the data samples already being ordered into an array (raw image patches, for example), while CG model opens this modeling strategy to a much wider set of data types.

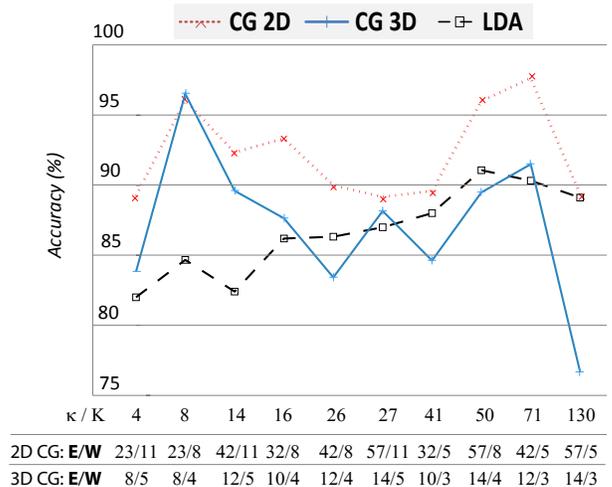

Figure 7: Microarray classification accuracy (%) for the prostate cancer dataset for various values of the capacity $\kappa$, and the number of LDA topics. The bottom of the table shows the actual widths of the square(cube) windows (**W**) and counting grids (**E**). All CGs are of uniform dimensions.

# References

U. Alon, N. Barkai, D. A. Notterman, K. Gishdagger, S. Ybarradagger, D. Mackdagger, and A. J. Levine. Broad patterns of gene expression revealed by clustering analysis of tumor and normal colon tissues probed by oligonucleotide arrays. *Proceedings of the National Academy of Sciences of the United States of America*, 96(12):6745–6750, June 1999.

A. Banerjee and S. Basu. Topic Models over Text Streams: A Study of Batch and Online Unsupervised Learning. In *Proceedings of SDM*, 2007.

D. Blei, A. Ng, and M. Jordan. Latent Dirichlet Allocation. *Journal of Machine Learning Research*, 3: 993–1022, January 2003.

A. Bosch, A. Zissermann, and X. Munoz. Scene classification via plsa. In *Proceedings of ECCV*, 2006.

S. Deerwester, S. Dumais, G. Furnas, T. Landauer, and R. Harshman. Indexing by latent semantic analysis. *Journal of the American Society for Information Science*, 41(6):391–407, 1990.

S. Dhanasekaran. Delineation of prognostic biomarkers in prostate cancer. *Nature*, 412(6849):822–826, 2001.

T. Hertz et al. Mapping the Landscape of Host-Pathogen Coevolution: HLA Class I Binding and Its Relationship with Evolutionary Conservation in Human and Viral Proteins. *Journal of Virology*, 85 (85):1310–1321, 2010.

T. Jaakkola and D. Haussler. Exploiting generative models in discriminative classifiers. In *In Advances in Neural Information Processing Systems 11*, 1998.

N. Jojic, B. Frey, and A. Kannan. Epitomic analysis of appearance and shape. In *Proceedings of ICCV*, 2003.

N. Jojic, A. Perina, and V. Murino. Structural epitome: a way to summarize one's visual experience. In *Advances in Neural Information Processing Systems 23*, 2010.

P. Kiepiela et al. Dominant influence of HLA-B in mediating the potential co-evolution of HIV and HLA. *Nature*, 432(85):769–775, 2004.

F Li and P. Perona. A Bayesian Hierarchical Model for Learning Natural Scene Categories. In *Proceedings of CVPR*, 2005.

K. Mardia and P. Jupp. *Directional Statistics*. John Wiley and Sons Ltd., 2000.

C. Moore et al. Evidence of HIV-1 Adaptation to HLA-Restricted Immune Responses at a Population Level. *Science*, 296(5572):436–442, 2002.

K. Ni, A. Kannan, A. Criminisi, and J. Winn. Epitomic location recognition. In *Proceedings of CVPR*, 2008.

A. Oliva and A. Torralba. Modeling the Shape of the Scene: A Holistic Representation of the Spatial Envelope. *International Journal of Computer Vision*, 42(3):145–175, 2001.

H. Peng, F. Long, and C. Ding. Feature selection based on mutual information: Criteria of max-dependency, max-relevance and min-redundancy. *IEEE Transactions on Pattern Analysis and Machine Intelligence*, 27(8):1226–1238, 2005.

A. Perina, M. Cristani, U. Castellani, V. Murino, and N. Jojic. Free energy score space. In *Advances in Neural Information Processing Systems 22*, 2009.

A. Perina and N. Jojic. Image analysis by counting on a grid. In *Proceedings of CVPR*, 2011.

A. Perina, P. Lovato, V. Murino, and M. Bicego. Biologically-aware latent dirichlet allocation (balda) for the classification of expression microarray. In *Proceedings of the 5th IAPR international conference on Pattern recognition in bioinformatics*, PRIB'10, pages 230–241, 2010.

S. Pomeroy. Prediction of central nervous system embryonal tumour outcome based on gene expression. *Nature*, 415(6870):436–442, 2002.

J. Reisinger, A. Waters, B. Silverthorn, and R. Mooney. Spherical topic models. In *Proceedings of ICML*, 2010.

S. Rogers, M. Girolami, C. Campbell, and R. Breitling. The latent process decomposition of cdna microarray data sets. *IEEE/ACM Transactions on Computational Biology and Bioinformatics*, 2:143–156, 2005.

G.G. Towell, J. Shavlik, and M. Noordewier. Refinement of approximate domain theories by knowledge-based neural networks. In *In Proceedings of the Eighth National Conference on Artificial Intelligence*, 1990.